\pdfoutput=1
\documentclass[sigconf,natbib=true,nonacm]{acmart}
\usepackage{ragged2e}
\newcommand{\lamp}{\textsc{Lamp}}
\newcommand{\mixmtp}{Mix-MTP}
\newcommand{\catalog}{\mathcal{I}}

\newcommand{\slotstate}[1]{\mathbf{h}_{#1}}

\newcommand{\btrain}{\beta_{\mathrm{train}}}
\newcommand{\beval}{\beta_{\mathrm{eval}}}

\newcommand{\lampArxivAssets}{}
\renewcommand\footnotetextcopyrightpermission[1]{}

\makeatletter
\if@ACM@natbib
  \@expandtwoargs\@removeelement{natbib=true}{\@unusedoptionlist}\@unusedoptionlist
\fi
\newcommand{\numberedaffiliation}[2]{#2}
\newcommand{\hkustaffiliation}{\numberedaffiliation{1}{%
  \department{Department of Mathematics}%
  \institution{The Hong Kong University of Science and Technology}%
  \city{Hong Kong SAR}\country{China}}}
\newcommand{\tripaffiliation}{\numberedaffiliation{2}{%
  \institution{Trip.com Group}\city{Shanghai}\country{China}}}
\renewcommand{\@mkauthors}{%
  \begingroup
  \count@=0\relax
  \let\and\relax
  \def\@author##1{\def\gridname{##1}}%
  \def\email##1##2{\def\gridemail{##2}}%
  \def\affiliation##1##2{##2}%
  \def\numberedaffiliation##1##2{%
    \begin{minipage}[t]{0.195\textwidth}%
      \centering\large\gridname\textsuperscript{##1}\par
      \normalsize\href{mailto:\gridemail}{\gridemail}\par
    \end{minipage}%
    \advance\count@ by 1\relax
    \ifnum\count@=5\par\smallskip\fi
    \ifnum\count@=9\par\smallskip\fi}%
  \global\setbox\mktitle@bx=\vbox{%
    \unvbox\mktitle@bx
    \par\medskip
    \centering\noindent\addresses
    \def\department##1{##1, }%
    \def\institution##1{##1, }%
    \def\city##1{##1, }%
    \def\country##1{##1}%
    \def\numberedaffiliation##1##2{\textsuperscript{##1}\,##2\par}%
    \normalsize\hkustaffiliation\tripaffiliation
    \medskip}%
  \endgroup}
\makeatother
\begin{document}
\title[LAMP: Language-Aligned Multi-Intent Parallel Decoding]{LAMP: Language-Aligned Multi-Intent Parallel Decoding of Semantic IDs for Generative Recommendation}
\author{Yixuan Xu}
\email{yxugr@connect.ust.hk}
\affiliation{\hkustaffiliation}
\author{Yuheng Chen}
\email{ychenlp@connect.ust.hk}
\affiliation{\hkustaffiliation}
\author{Fan Zhang}
\email{mafzhang@ust.hk}
\affiliation{\hkustaffiliation}
\author{Hailin Liu}
\email{hliuee@connect.ust.hk}
\affiliation{\hkustaffiliation}
\author{Zhixi Xiong}
\email{zxiongal@connect.ust.hk}
\affiliation{\hkustaffiliation}
\author{Chao Yang}
\email{cyang3@trip.com}
\affiliation{\tripaffiliation}
\author{Yang Xiang}
\email{yxiang@trip.com}
\affiliation{\tripaffiliation}
\author{Chenji Pan}
\email{chenjipan@trip.com}
\affiliation{\tripaffiliation}
\author{Can Yang}
\email{macyang@ust.hk}
\affiliation{\hkustaffiliation}
\renewcommand{\shortauthors}{Xu et al.}
\begin{abstract}
Generative recommenders retrieve next items by predicting discrete semantic IDs (SIDs), yet existing decoders face an acute tension among index resolution, decoding efficiency, and multi-intent expressiveness. Autoregressive decoders serialize long codes across multiple backbone passes and risk pruning valid targets via beam search. Parallel decoders score flat codes in one pass, but predicting a single set of position-wise marginals breaks intra-item codeword coupling and can tie unsupported ``ghost'' tuples with plausible purchases. We introduce \textbf{\lamp{}} (\emph{Language-Aligned Multi-Intent Parallel Decoding}), a single-pass generative recommender that preserves codeword coupling by scoring each candidate's complete ID under multiple learned intent slots before pooling via LogSumExp. During training, a soft winner-take-all mixture loss concentrates gradients on slots that better explain the observed target, while two training-only bidirectional language objectives---next-title prediction and item-text-to-SID alignment through the shared code head---anchor slot specialization in discrete code space at zero inference cost. Across three Amazon domains and three seeds, \lamp{} improves NDCG@10 by 25.0--29.9\% over RPG; on the same Qwen3-0.6B backbone, it outperforms MiniOneRec-SFT-AR by 89.2--127.9\% while cutting single-request p95 latency by $4.7\times$--$11.9\times$ and raising batched throughput by $13.4\times$--$78.3\times$. Controlled ablations confirm that bidirectional grounding significantly widens the four-slot advantage over single-readout baselines across all three domains, while item-text alignment yields its largest lifts on low-frequency targets ($+37.5$--$71.8\%$). \lamp{} further transfers to training-free SimHash IDs, improves on Toys from 0.6B to 4B parameters, and ranks a resampled one-million-item catalog in 24.0\,ms at p95.
\end{abstract}
\maketitle
\raggedbottom
\setlength{\emergencystretch}{1em}

\section{Introduction}
\label{sec:intro}
Generative recommenders retrieve the next item by predicting its discrete semantic ID (SID) from a user's interaction history~\cite{tiger,minionerec}. Representing items as tuples of codewords lets rarely purchased items share parameters with popular ones, but it creates a trade-off between index resolution and decoding depth.
\begin{figure}[t]
\centering
\includegraphics[width=\columnwidth]{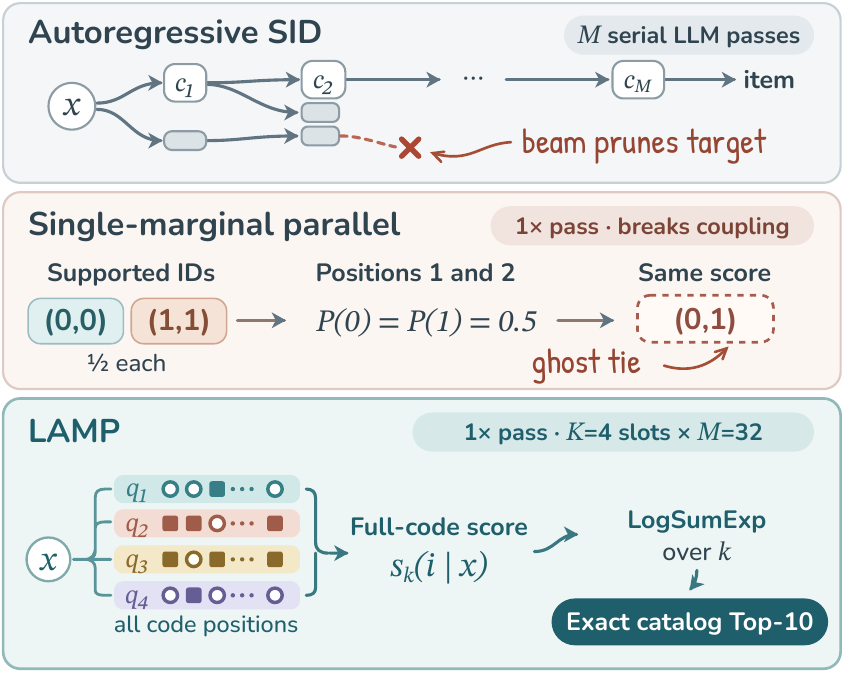}
\caption{A legal-prefix beam can prune the target early. Independent code marginals can tie an unsupported pair $(0,1)$ with both plausible targets. \lamp{} scores each item's complete ID under each slot before pooling the scores for exact catalog ranking.}
\Description{Three panels show a pruned branch in an autoregressive prefix tree, a tie induced by independent marginals for equally likely targets (0,0) and (1,1), and four intent slots that score each full item code before LogSumExp aggregation.}
\label{fig:overview}
\end{figure}

Short identifiers force distinct items to share an exact code and therefore receive identical scores. Lengthening an autoregressive code resolves these collisions, yet each extra position adds another serial backbone pass and gives prefix-constrained beam search another chance to prune a valid target before its distinguishing codeword is reached.

Parallel decoders break this serial bottleneck by predicting all positions of a flat code in a single forward pass~\cite{rpg,setrec,flash}. Predicting one set of position-wise marginals, however, treats code positions as conditionally independent given the history. When a user's history supports more than one plausible next purchase, this unimodal factorization breaks intra-item codeword coupling. Suppose a history equally favors two items with binary IDs $(0,0)$ and $(1,1)$ (Figure~\ref{fig:overview}). Each position has a uniform $0.5$ marginal, so multiplying the marginals assigns the unsupported ``ghost'' tuple $(0,1)$ the exact same score as either valid target. If $(0,1)$ appears in the catalog, even exact full-catalog search cannot break the tie.

A natural fix is to maintain multiple parallel readouts that score each candidate's complete ID under competing interpretations before pooling across intents. In discrete SID space, however, adding intent slots encounters a second, coupled obstacle: quantized codewords are opaque integers whose meaning is local to a codebook rather than aligned with a pretrained language model. While product titles explicitly name the attributes that separate competing user intents, next-item supervision alone supplies sparse updates for the exact code combinations of infrequent items. Without semantic anchoring in code space, multi-slot decoding realizes only a fraction of its potential---yielding modest gains on Toys and virtually no improvement on Beauty or Sports (Section~\ref{sec:exp-ablation}).

We address both problems jointly with \lamp{} (\emph{Language-Aligned Multi-Intent Parallel Decoding}), a single-pass generative recommender (Figure~\ref{fig:overview}). Its \mixmtp{} decoder appends $K$ learned intent slots to the user's ordered title history. Under causal attention, each slot predicts all $M$ positions of a long, flat SID through a shared classifier head $W$. The decoder scores every catalog item's complete code under each slot and pools the slot scores via LogSumExp for exact full-catalog ranking. During training, a soft winner-take-all mixture loss assigns more gradient weight to slots that better explain the observed target, while a batch-level regularizer encourages balanced average slot use.

To anchor both sides of this shared scoring space without slowing down inference, \lamp{} adds bidirectional language grounding during training only. In the forward direction, the backbone's tied language head predicts the next item's title from the history and intent slots, aligning slot states with native vocabulary semantics. In the reverse direction, a dedicated item readout maps each training-visible item's title and fine-grained category directly to its SID through the same head $W$, providing dense text-to-code supervision even for rarely purchased items. Because both objectives are dropped after training, inference still requires only a single backbone pass and an exact catalog lookup.

Our main contributions are threefold:
\begin{itemize}
\item \textbf{Coupling-preserving multi-intent parallel decoding.} We formalize how unimodal parallel SID decoders suffer from rank-one factorization and unsupported ``ghost'' ties, and introduce \lamp{} (\mixmtp{}), a single-pass recommender that scores complete item codes under multiple causal intent slots before LogSumExp pooling. Pairing soft winner-take-all gradient routing with training-only bidirectional language grounding anchors slot specialization in discrete code space at zero inference cost.
\item \textbf{Mechanistic evidence and tokenizer generality.} Controlled ablations show that bidirectional grounding significantly widens the four-slot advantage over the corresponding single readout across all three domains (request-paired 95\% CIs above zero), while replacing mean-slot CE with our mixture loss under the identical four-slot setup on Toys raises NDCG@10 by $+0.832$ (5.703 to 6.535). Item-text alignment yields its largest gains on low-frequency targets (1--5 training interactions, $+37.5$--$71.8\%$), and the same decoder transfers directly to training-free SimHash IDs, performing comparably to fitted OPQ on Toys.
\item \textbf{Accuracy, scaling, and serving efficiency.} Across three Amazon domains and three seeds under a unified full-catalog protocol, \lamp{} improves NDCG@10 by 25.0--29.9\% over RPG; on the same Qwen3-0.6B backbone, it outperforms MiniOneRec-SFT-AR by 89.2--127.9\% while cutting single-request p95 latency by $4.7\times$--$11.9\times$ and raising batched throughput by $13.4\times$--$78.3\times$. Accuracy improves from 0.6B to 4B parameters on Toys, and the 0.6B model ranks a resampled one-million-item catalog in 24.0\,ms at p95.
\end{itemize}

\begin{figure*}[t]
\centering
\includegraphics[width=\textwidth]{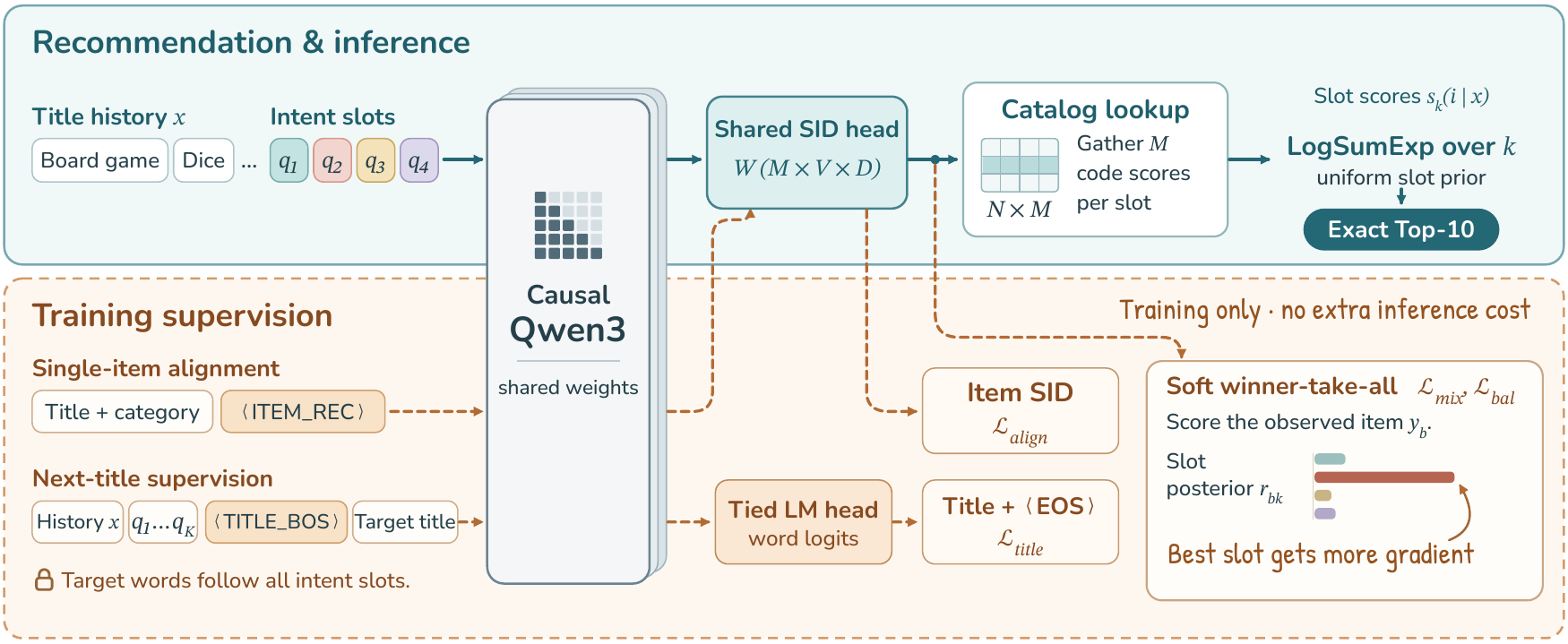}
\caption{In one backbone pass, \lamp{} scores every catalog item under each intent slot and combines the scores across slots. During training, a single-item readout predicts its SID through the same head $W$. The tied language head predicts the target title. Training responsibilities depend on the observed item's code.}
\Description{One causal Qwen3 backbone encodes all three prompts. At inference, the SID head and catalog table produce per-slot scores, which LogSumExp combines for exact ranking. During training the item readout feeds the shared SID head, and the title-start and subsequent states feed the tied language head. A separate training branch weights each slot's gradient using its score on the observed next item.}
\label{fig:architecture}
\end{figure*}

\section{Preliminaries}
\label{sec:prelim}

Let $\catalog$ be a catalog of $N$ items and $x=(i_1,\ldots,i_t)$ a user's chronologically ordered interaction history. Given $x$, the task is to rank the observed next item $y=i_{t+1}$ against every item in $\catalog$, including those already in the history. The recommender receives the titles of the items in $x$ and returns a ranked list of candidate items.

\paragraph{Semantic IDs}
Building on discrete vector quantization~\cite{vqvae}, a semantic ID represents an item by $M$ discrete codewords~\cite{tiger}. Optimized product quantization (OPQ)~\cite{opq} rotates the item embedding, partitions it into $M$ blocks, and quantizes each block with a separate codebook of size $V$. The resulting ID is
\begin{equation}
 \mathbf c(i)=(c_1(i),\ldots,c_M(i)),\qquad c_m(i)\in\{0,\ldots,V-1\}.
 \label{eq:lamp-code}
\end{equation}
Each position identifies a fixed subspace, so $c_m(i)$ can be predicted without first generating $c_{<m}(i)$~\cite{rpg}. Items with the same tuple receive the same score. For a bucket $\mathcal B(\mathbf c)=\{i\in\catalog:\mathbf c(i)=\mathbf c\}$, we measure the share of catalog items affected by collisions as
\[
 \frac{1}{N}\sum_{i\in\catalog}
    \mathbf 1\!\left[|\mathcal B(\mathbf c(i))|>1\right].
\]
All members of a bucket remain eligible at evaluation under a fixed item-level tie rule.

\paragraph{Code length and retrieval depth}
Product~\cite{pq} and residual~\cite{rqvae,tiger} codes each allocate $M\log_2 V$ bits per item. A product code quantizes $M$ fixed subspaces independently, whereas each residual stage quantizes the error left by the preceding stages. Autoregressive search conditions on each surviving prefix and takes $M$ sequential prediction steps. A parallel decoder can instead extend the code while encoding the history once. Section~\ref{sec:method} shows how to score a complete tuple under multiple plausible interpretations of that history.

\section{Language-Aligned Multi-Intent Parallel Decoding}
\label{sec:method}

\lamp{} adapts a causal language model to rank items under multiple interpretations of a user's history (Figure~\ref{fig:architecture}). Its multi-token prediction decoder, \mixmtp{}, predicts all code positions under each interpretation in a single forward pass. Two language objectives connect this predictor to the text of the items it ranks.

\subsection{Multi-Intent Parallel Code Prediction}
Let $x_{\rm text}$ be the embedded title sequence in purchase order. The recommendation input is $[x_{\rm text};\,\mathbf q_1,\ldots,\mathbf q_K]$, where the $\mathbf q_k$ are learned vectors. Under the causal mask, slot $k$ attends to the history, itself, and preceding slots $\mathbf q_{<k}$. Permuting the slots changes which earlier states each one can attend to, breaking permutation symmetry. These distinct contexts allow slots with a shared output head to specialize without an orthogonality constraint.

{\interlinepenalty=10000\relax For subspace $m$, the final state $\slotstate{k}\in\mathbb R^D$ and the bias-free head $W\in\mathbb R^{M\times V\times D}$ assign codeword $v$ the probability
\begin{equation}
 p_{km}(v\mid x)=
 \frac{\exp(W_{mv}^{\mathsf T}\slotstate{k})}
      {\sum_{u=0}^{V-1}\exp(W_{mu}^{\mathsf T}\slotstate{k})}.
 \label{eq:lamp-head}
\end{equation}\par}
For candidate $i$, slot $k$ scores its complete identifier by gathering the log probabilities of its $M$ codewords,
\begin{equation}
 s_k(i\mid x)=\frac{1}{M}\sum_{m=1}^M\log p_{km}\big(c_m(i)\mid x\big).
 \label{eq:lamp-slot}
\end{equation}
Averaging across positions gives the log geometric mean of the codeword probabilities, which keeps scores on a comparable scale across code lengths. The classifiers $W_m$ are specific to code positions and shared across slots. A single matrix multiplication of the $K$ slot states against the $MV$ rows of $W$, followed by normalization within each position, thus yields every distribution in Eq.~\eqref{eq:lamp-head}.

\subsection{Soft Winner-Take-All Mixture Training}
The unimodal factorization $P(\mathbf c\mid x)=\prod_m p_m(c_m\mid x)$ defines a rank-one tensor over code tuples. In the binary example of Section~\ref{sec:intro}, uniform marginals produce a ``ghost'' tie between the unsupported tuple $(0,1)$ and the two targets $(0,0)$ and $(1,1)$. A mixture $\sum_k\pi_k P_k$ can preserve intra-item codeword coupling because each slot scores a complete tuple under its own distribution $P_k(\mathbf c\mid x)=\prod_m p_{km}(c_m\mid x)$ before pooling. In \lamp{}, Eq.~\eqref{eq:lamp-slot} gives $\exp(\beta s_k)=P_k^{\beta/M}$ for each candidate. At $\beta=M$, pooling these terms with weights $\pi_k$ gives the normalized mixture over code tuples. For long codes such as $M=32$, choosing $0<\btrain<M$ tempers each complete-code likelihood before mixing, reducing the risk that one slot takes nearly all the responsibility early in training.

The binary example makes the resulting separation explicit. Suppose one slot assigns probability $1-\epsilon$ to zero at both positions and the other assigns the same probability to one, with $0<\epsilon<1/2$. Set $a=(1-\epsilon)^{\beta/2}$ and $d=\epsilon^{\beta/2}$. With equal slot weights, the pooled scores before the logarithm are $(a^2+d^2)/2$ for either supported tuple and $ad$ for the crossed tuple. Their difference,
\begin{equation}
 \tfrac12(a^2+d^2)-ad=\tfrac12(a-d)^2>0,
 \label{eq:lamp-binary-gap}
\end{equation}
holds at every positive scale, although the two position marginals pooled separately would still be uniform.

For a recommendation microbatch of $B$ examples, write $s_{bk}=s_k(y_b\mid x_b)$ for slot $k$'s score on the observed target. With a uniform prior $\pi_k=1/K$ and $\btrain>0$, the mixture loss is
\begin{equation}
 \mathcal L_{\rm mix}
 =-\frac{1}{B\btrain}\sum_{b=1}^{B}
   \log\!\left(\sum_{k=1}^K\pi_k\exp(\btrain s_{bk})\right).
 \label{eq:lamp-mix}
\end{equation}
At $K=1$, division by $\btrain$ recovers mean code-position cross-entropy (CE). For $K>1$, let $\bar r_k=\frac{1}{B}\sum_{b=1}^{B}r_{bk}$ be the microbatch mean of slot $k$'s posterior responsibility for the observed code. The responsibility and the loss that balances its mean against the uniform prior $U_K$ are
\begin{equation}
 \begin{aligned}
 r_{bk}&=\frac{\pi_k\exp(\btrain s_{bk})}
               {\sum_j\pi_j\exp(\btrain s_{bj})},\\
 \mathcal L_{\rm bal}&=\mathrm{KL}(\bar{\mathbf r}\,\|\,U_K)
   =\sum_k\bar r_k\log(K\bar r_k).
 \end{aligned}
 \label{eq:lamp-balance}
\end{equation}
The derivative $\partial\mathcal L_{\rm mix}/\partial s_{bk}=-r_{bk}/B$ routes supervision for every code position of a target through the same responsibility. For request $b$ with slot state $\slotstate{bk}$, write $z_{bkmv}=W_{mv}^{\mathsf T}\slotstate{bk}$ and $p_{bkmv}=p_{km}(v\mid x_b)$. Then
\begin{equation}
 \frac{\partial\mathcal L_{\rm mix}}{\partial z_{bkmv}}
 =\frac{r_{bk}}{BM}\big(p_{bkmv}-\mathbf 1\{v=c_m(y_b)\}\big).
 \label{eq:lamp-logit-gradient}
\end{equation}
At the shared head this gives the explicit state gradient
\begin{equation}
 \nabla_{\slotstate{bk}}\mathcal L_{\rm mix}
 =\sum_{m,v} W_{mv}\,\frac{\partial\mathcal L_{\rm mix}}{\partial z_{bkmv}}.
 \label{eq:lamp-state-gradient}
\end{equation}
A slot that explains the complete item consequently receives a larger share of the update at every position.

The inverse temperature controls this division of credit. As $\btrain\to0^+$, $r_{bk}\to1/K$ and Eq.~\eqref{eq:lamp-mix} tends to $-(BK)^{-1}\sum_{b,k}s_{bk}$, the mean of the individual slot CEs. Every slot then learns from every target with equal weight, pushing slots toward redundant predictions. Section~\ref{sec:exp-ablation} shows the ranking cost of equal weighting with the inference rule held fixed. As $\btrain\to\infty$, the loss approaches $-B^{-1}\sum_b\max_k s_{bk}$. Responsibility concentrates on each target's best-scoring slot, splitting equally among tied winners. Slots that never win can then receive vanishingly small mixture gradients. Finite temperature gives competing slots credit, while $\mathcal L_{\rm bal}$ favors balanced usage across the microbatch. Identical slot predictions also give $\mathcal L_{\rm bal}=0$, so Section~\ref{sec:exp-ablation} measures prediction diversity on Beauty and Sports through cross-slot Top-10 overlap and per-position Jensen--Shannon divergence.

\subsection{Training-Only Bidirectional Language Grounding}
\paragraph{History to target title}
A learned title-start vector follows the intent slots during training. Beginning at that position, the native language head predicts the next item's title and end-of-sequence token by teacher forcing. Let $t_{b,1:L_b}$ include the title tokens and end token for request $b$, and let $u_b$ comprise the history, intent slots, and title-start position. The language head shares its output projection with the input embeddings and minimizes mean cross-entropy over labeled tokens,
\begin{equation}
 \mathcal L_{\rm title} =
 -\frac{1}{\sum_b L_b}\sum_b\sum_{j=1}^{L_b}
 \log p_{\rm LM}(t_{b,j}\mid u_b,t_{b,<j}).
 \label{eq:lamp-title}
\end{equation}
Title prediction uses the history and the intent slots. Causal attention keeps the target words outside the context of every recommendation slot.

\paragraph{Item text to SID}
In the reverse direction, the same backbone reads a training-visible item's title and fine-grained category, and a dedicated item slot predicts its SID through $W$. To ensure unambiguous supervision, we filter out tokenized prompts that map to multiple distinct SIDs. For an alignment minibatch $\mathcal A$, write $a_{im}$ for the SID head's categorical distribution at position $m$ when it reads item $i$ in isolation. Its cross-entropy is
\begin{equation}
 \mathcal L_{\rm align} =
 -\frac{1}{|\mathcal{A}|M}\sum_{i\in\mathcal{A}}\sum_{m=1}^M
         \log a_{im}\big(c_m(i)\big).
 \label{eq:lamp-align}
\end{equation}
The geometry of the alignment update follows directly from the shared linear head. If $\mathbf g_i$ is the item slot's final state, then for position $m$ and codeword $v$,
\begin{equation}
 \frac{\partial\mathcal L_{\rm align}}{\partial W_{mv}}
 =\frac{1}{|\mathcal A|M}\sum_{i\in\mathcal A}
 \big(a_{im}(v)-\mathbf 1\{c_m(i)=v\}\big)\mathbf g_i .
 \label{eq:lamp-align-gradient}
\end{equation}
Under gradient descent, an item contributes a step toward its text state for each target row, and away from that state for competing rows in proportion to their predicted probabilities. Under next-item supervision alone, an item directly updates its target rows $W_{m,c_m(i)}$ each time it appears as a training target, leaving rare items with few updates for their specific code combinations. Alignment samples uniformly over eligible training-visible items, independently of their interaction counts. Each sampled item's text thus trains all $M$ target rows of $W$, including those later queried by a history slot. Title prediction acts on the history side of the dot product $W_{mv}^{\mathsf T}\slotstate{k}$. Gradients from the native vocabulary pass through the shared backbone to the history and slots. Together, the two paths align each history interpretation with the codewords supported by its language. Consistent with this account, language grounding increases the measured gain from multiple slots in all three domains (Section~\ref{sec:exp-ablation}).

Only recommendation examples contribute to the mixture and balancing terms. Each auxiliary loss has its own mean, giving the joint objective
\begin{equation}
 \mathcal L=\mathcal L_{\rm mix}+\lambda_{\rm bal}\mathcal L_{\rm bal}
           +\lambda_{\rm title}\mathcal L_{\rm title}
           +\lambda_{\rm align}\mathcal L_{\rm align}.
 \label{eq:lamp-total}
\end{equation}

\subsection{Single-Pass Exact Full-Catalog Ranking}
At inference, the prompt ends at the last intent slot. A single backbone pass yields the slot states for catalog ranking, and the two auxiliary tasks add no computation. With evaluation scale $\beval>0$, the catalog score is
\begin{equation}
 S(i\mid x)=\log\!\left(\sum_{k=1}^K\pi_k
                  \exp\!\left(\beval s_k(i\mid x)\right)\right).
 \label{eq:lamp-rank}
\end{equation}
The $N\times M$ catalog table indexes the $K\times M\times V$ log probabilities produced for a request. For each item, lookup yields one log probability per position and slot, so Eq.~\eqref{eq:lamp-rank} pools the slot means directly. Dropping the positive factor $1/\beval$ preserves the ordering of items.

Because all $KM$ distributions come from a single backbone pass, a longer code increases the $O(KMVD)$ cost of the head and the $O(NKM)$ cost of catalog scoring while leaving decoding depth fixed. An $O(N\log N)$ sort of all scores returns the exact Top-10, breaking ties by ascending item ID. Autoregressive beam search requires $M$ successive prefix-conditioned predictions even with a key--value cache.

Discrete codes also keep the catalog index compact. For one million items, 32 one-byte codes per item occupy 32\,MB when packed, compared with 1,024\,MB for one 512-dimensional FP16 vector per item.

% Generated by scripts/phase1_tables.py
\begin{table*}[t]
\caption{Test Recall (R) and NDCG (N), scaled by 100. The upper block quotes RPG Table~2 under its published protocol for context. All direct comparisons and lifts use the lower block's full-catalog protocol, retaining repeated interactions and previously visited candidates. Sample SDs ($\pm$) appear beneath the means. Bold and underline mark the best and second-best means within this block. Lifts use unrounded means. The last row gives MiniOneRec-SFT-AR $\to$ \lamp{} p95 (speedup) from tokenized history to CPU Top-10 (seed 42, batch 1, exclusive RTX 6000D). Domain-order settings are beams 20/20/100 and $\beta=4/2/2$, respectively.}
\label{tab:main}
\centering\small
\renewcommand{\arraystretch}{1.0}
\setlength{\tabcolsep}{1.75pt}
\begin{tabular*}{\linewidth}{@{\extracolsep{\fill}}l*{4}{r}@{\hspace{13pt}}*{4}{r}@{\hspace{13pt}}*{4}{r}@{}}
\toprule
 & \multicolumn{4}{c}{\textbf{Toys}} & \multicolumn{4}{c}{\textbf{Beauty}} & \multicolumn{4}{c}{\textbf{Sports}} \\
\cmidrule(lr){2-5}\cmidrule(lr){6-9}\cmidrule(lr){10-13}
Method & R@5 & R@10 & N@5 & N@10 & R@5 & R@10 & N@5 & N@10 & R@5 & R@10 & N@5 & N@10 \\
\midrule
\multicolumn{13}{@{}l}{\textit{Published baselines (from RPG, Table~2~\cite{rpg})}} \\
Caser & 1.66 & 2.70 & 1.07 & 1.41 & 2.05 & 3.47 & 1.31 & 1.76 & 1.16 & 1.94 & 0.72 & 0.97 \\
GRU4Rec & 0.97 & 1.76 & 0.59 & 0.84 & 1.64 & 2.83 & 0.99 & 1.37 & 1.29 & 2.04 & 0.86 & 1.10 \\
HGN & 3.21 & 4.97 & 2.21 & 2.77 & 3.25 & 5.12 & 2.06 & 2.66 & 1.89 & 3.13 & 1.20 & 1.59 \\
BERT4Rec & 1.16 & 2.03 & 0.71 & 0.99 & 2.03 & 3.47 & 1.24 & 1.70 & 1.15 & 1.91 & 0.75 & 0.99 \\
SASRec & 4.63 & 6.75 & 3.06 & 3.74 & 3.87 & 6.05 & 2.49 & 3.18 & 2.33 & 3.50 & 1.54 & 1.92 \\
FDSA & 2.28 & 3.81 & 1.40 & 1.89 & 2.67 & 4.07 & 1.63 & 2.08 & 1.82 & 2.88 & 1.22 & 1.56 \\
S$^3$-Rec & 4.43 & 7.00 & 2.94 & 3.76 & 3.87 & 6.47 & 2.44 & 3.27 & 2.51 & 3.85 & 1.61 & 2.04 \\
RecJPQ & 3.31 & 4.84 & 1.82 & 2.31 & 3.11 & 4.82 & 1.67 & 2.22 & 1.41 & 2.20 & 0.76 & 1.02 \\
VQ-Rec & 4.97 & 7.37 & 3.46 & 4.23 & 4.57 & 6.64 & 3.17 & 3.83 & 2.08 & 3.00 & 1.44 & 1.73 \\
TIGER & 5.21 & 7.12 & 3.71 & 4.32 & 4.54 & 6.48 & 3.21 & 3.84 & 2.64 & 4.00 & 1.81 & 2.25 \\
HSTU & 4.33 & 6.69 & 2.81 & 3.57 & 4.69 & 7.04 & 3.14 & 3.89 & 2.58 & 4.14 & 1.65 & 2.15 \\
\midrule
\multicolumn{13}{@{}l}{\textit{Controlled full-catalog evaluation (seeds 42--44)}} \\
RPG & \underline{6.031} & \underline{9.020} & \underline{4.038} & \underline{4.998} & \underline{5.454} & \underline{8.024} & \underline{3.797} & \underline{4.625} & \underline{3.069} & \underline{4.463} & \underline{2.138} & \underline{2.585} \\[-2pt]
 & {\scriptsize $\pm\,0.049$} & {\scriptsize $\pm\,0.057$} & {\scriptsize $\pm\,0.042$} & {\scriptsize $\pm\,0.046$} & {\scriptsize $\pm\,0.082$} & {\scriptsize $\pm\,0.054$} & {\scriptsize $\pm\,0.049$} & {\scriptsize $\pm\,0.042$} & {\scriptsize $\pm\,0.058$} & {\scriptsize $\pm\,0.097$} & {\scriptsize $\pm\,0.041$} & {\scriptsize $\pm\,0.028$} \\
\addlinespace[3pt]
MiniOneRec-SFT-AR & 3.520 & 5.512 & 2.206 & 2.848 & 3.561 & 5.657 & 2.279 & 2.953 & 2.097 & 3.367 & 1.366 & 1.774 \\[-2pt]
 & {\scriptsize $\pm\,0.131$} & {\scriptsize $\pm\,0.240$} & {\scriptsize $\pm\,0.106$} & {\scriptsize $\pm\,0.139$} & {\scriptsize $\pm\,0.195$} & {\scriptsize $\pm\,0.344$} & {\scriptsize $\pm\,0.091$} & {\scriptsize $\pm\,0.142$} & {\scriptsize $\pm\,0.015$} & {\scriptsize $\pm\,0.020$} & {\scriptsize $\pm\,0.026$} & {\scriptsize $\pm\,0.025$} \\
\addlinespace[3pt]
\textbf{\lamp{}} & \textbf{7.631} & \textbf{10.602} & \textbf{5.531} & \textbf{6.490} & \textbf{6.775} & \textbf{9.499} & \textbf{4.909} & \textbf{5.782} & \textbf{3.992} & \textbf{5.814} & \textbf{2.771} & \textbf{3.357} \\[-2pt]
 & {\scriptsize $\pm\,0.037$} & {\scriptsize $\pm\,0.209$} & {\scriptsize $\pm\,0.048$} & {\scriptsize $\pm\,0.095$} & {\scriptsize $\pm\,0.045$} & {\scriptsize $\pm\,0.084$} & {\scriptsize $\pm\,0.086$} & {\scriptsize $\pm\,0.102$} & {\scriptsize $\pm\,0.128$} & {\scriptsize $\pm\,0.087$} & {\scriptsize $\pm\,0.072$} & {\scriptsize $\pm\,0.064$} \\
\midrule
\emph{Lift over RPG (\%)} & +26.54 & +17.53 & +36.99 & +29.86 & +24.21 & +18.39 & +29.31 & +25.04 & +30.09 & +30.28 & +29.63 & +29.85 \\
\emph{Lift over MiniOneRec-SFT-AR (\%)} & +116.78 & +92.34 & +150.71 & +127.92 & +90.25 & +67.93 & +115.40 & +95.80 & +90.40 & +72.66 & +102.87 & +89.17 \\
\midrule
p95 (ms) & \multicolumn{4}{c}{107.76 $\to$ \textbf{22.97} (\textbf{4.7\boldmath$\times$})} & \multicolumn{4}{c}{104.93 $\to$ \textbf{21.47} (\textbf{4.9\boldmath$\times$})} & \multicolumn{4}{c}{244.34 $\to$ \textbf{20.52} (\textbf{11.9\boldmath$\times$})} \\
\bottomrule
\end{tabular*}
\end{table*}

\section{Experiments}
\label{sec:experiments}
\begin{list}{}{%
  \settowidth{\labelwidth}{\textbf{RQ5:}}%
  \setlength{\labelsep}{0.45em}%
  \setlength{\leftmargin}{\dimexpr\labelwidth+\labelsep\relax}%
  \setlength{\topsep}{0pt}\setlength{\partopsep}{0pt}%
  \setlength{\itemsep}{0pt}\setlength{\parsep}{0pt}}
\item[\textbf{RQ1:}] How accurately does \lamp{} rank the next item against sequential and generative baselines?
\item[\textbf{RQ2:}] What do multiple slots and language grounding contribute, and does the decoder depend on a particular tokenizer?
\item[\textbf{RQ3:}] Which target items gain most from alignment?
\item[\textbf{RQ4:}] How sensitive is ranking to code length, slot count, and loss weights?
\item[\textbf{RQ5:}] How do backbone size, catalog size, and serving batch size affect accuracy and cost?
\end{list}

\subsection{Experimental Setup}
\paragraph{Data and evaluation}
We use three five-core domains from Amazon Reviews 2014~\cite{amazon14styles,amazon14trends}. In the order used throughout, Toys and Games, Beauty, and Sports and Outdoors contain 19,412, 22,363, and 35,598 users and 11,924, 12,101, and 18,357 items, respectively. Chronological leave-one-out splitting assigns each user's last interaction to test and the preceding one to validation, yielding one request per user in each held-out split and 109,361, 131,413, and 189,543 training prefix--target pairs from earlier interactions. At test time, the observed history includes the validation target. Training and retrieval use the most recent 20 available events, including repeats, and retrieval ranks the full catalog, including previously visited items. We expand shared codes to their member items, break score ties by ascending item ID, and report Recall and NDCG at 5 and 10 on a $100\times$ scale.

\paragraph{Baselines and controlled comparisons}
For historical context, the upper block of Table~\ref{tab:main} quotes RPG Table~2~\cite{rpg} under its published protocol. The published baselines include Caser~\cite{caser}, GRU4Rec~\cite{gru4rec}, HGN~\cite{hgn}, BERT4Rec~\cite{bert4rec}, SASRec~\cite{sasrec}, FDSA~\cite{fdsa}, S$^3$-Rec~\cite{s3rec}, RecJPQ~\cite{recjpq}, VQ-Rec~\cite{vqrec}, TIGER~\cite{tiger}, and HSTU~\cite{hstu}. All direct comparisons and lifts use the lower block, where RPG, MiniOneRec-SFT-AR, and \lamp{} rank the same full catalog for each request. All three build their IDs from the same 3,072-dimensional item vectors. Following RPG~\cite{rpg}, OpenAI's \texttt{text-embedding-3-large} encodes each item's concatenated title, price, brand, product features, category paths, and description, using the available fields. During training, \lamp{} also includes the fine-grained category in the single-item alignment prompt (Section~\ref{sec:method}). RPG and \lamp{} fit 512-dimensional PCA on the domain's static catalog and OPQ on training-visible items. RPG uses whitened PCA and 16, 32, and 16 code positions to predict product codes from past codes. \lamp{} uses non-whitened PCA and a separate OPQ32 in each domain.

For a matched supervised comparison on Qwen3-0.6B-Base, we evaluate MiniOneRec-SFT-AR, the supervised autoregressive stage of MiniOneRec~\cite{minionerec}. It predicts the next SID or title from past SIDs and translates between training items' titles and SIDs. Its RQ-VAE fits three 256-entry codebooks to training-visible vectors scaled by $\sqrt{3072}$, balances the first layer with Sinkhorn ($\epsilon=0.003$), and reassigns the last code within collision groups during full-catalog encoding. All seeds use the same index and constrained beam search. A separate \emph{RQ4-AR} control uses \lamp{}'s title history and Qwen3 backbone size, with a 256-way classifier for each of four residual $k$-means positions and beam width 50.

% Generated by scripts/phase1_tables.py
\begin{table*}[t]
\caption{Language grounding, readout, and tokenizer ablations ($100\times$ Test NDCG@10, seed 42, $\beta=4$ for mixtures). $I$ is the four-slot gain with both objectives minus the gain without grounding (request-paired 95\% CI). The ungrounded single-slot baseline reads the final prompt state, whereas the grounded single-slot variant appends one learned slot. Bold marks positive interactions and the best score in each domain (a) or panel (b,c).}
\label{tab:mechanism}
\centering\small
\renewcommand{\arraystretch}{1.16}
\setlength{\tabcolsep}{3.5pt}
\begin{tabular*}{\linewidth}{@{\extracolsep{\fill}}l rr@{\hspace{13pt}}rr@{\hspace{13pt}}rr@{\hspace{13pt}}rl@{}}
\toprule
\multicolumn{9}{@{}l}{\textbf{(a) Grounding $\times$ multi-slot synergy}} \\
 & \multicolumn{2}{c}{No grounding} & \multicolumn{2}{c}{One objective ($K=4$)} & \multicolumn{2}{c}{Both objectives} & \multicolumn{2}{c}{Interaction} \\
\cmidrule(lr){2-3}\cmidrule(lr){4-5}\cmidrule(lr){6-7}\cmidrule(lr){8-9}
Domain & $K=1$ & $K=4$ & Title & Align. & $K=1$ & $K=4$ & $I$ & Paired 95\% CI \\
\midrule
Toys & 5.374 & 5.827 & 5.888 & 6.388 & 5.664 & \textbf{6.535} & \textbf{+0.418} & [+0.176, +0.661] \\
Beauty & 4.888 & 4.863 & 5.193 & 5.604 & 5.025 & \textbf{5.809} & \textbf{+0.809} & [+0.568, +1.047] \\
Sports & 2.772 & 2.819 & 3.158 & 3.234 & 2.780 & \textbf{3.428} & \textbf{+0.601} & [+0.450, +0.757] \\
\bottomrule
\end{tabular*}\par\vspace{6pt}\noindent
\begin{minipage}[t]{0.48\textwidth}
\begin{tabular*}{\linewidth}{@{\extracolsep{\fill}}lr@{}}
\toprule
\multicolumn{2}{@{}l}{\textbf{(b) Readout \& loss ablation (Toys)}} \\
Readout on OPQ32 & NDCG@10 \\
\midrule
One slot, code CE & 5.664 \\
Four slots, last-slot CE & 5.823 \\
Four slots, mean slot CE & 5.703 \\
Four slots, mixture loss & \textbf{6.535} \\
\bottomrule
\end{tabular*}
\end{minipage}%
\hfill%
\begin{minipage}[t]{0.48\textwidth}
\begin{tabular*}{\linewidth}{@{\extracolsep{\fill}}lr@{}}
\toprule
\multicolumn{2}{@{}l}{\textbf{(c) Tokenizer generality (Toys)}} \\
Tokenizer with four-slot mixture & NDCG@10 \\
\midrule
Residual quantization, 4 codes & 4.992 \\
OPQ, 32 codes & 6.535 \\
SimHash, raw text vectors & 6.413 \\
SimHash, PCA text vectors & \textbf{6.614} \\
\bottomrule
\end{tabular*}
\end{minipage}%
\end{table*}

\paragraph{Implementation details}
Unless noted, \lamp{} builds on Qwen3-0.6B-Base~\cite{qwen3} with $(K,M,V)=(4,32,256)$ and $\btrain=4$. The loss weights $(\lambda_{\rm bal},\lambda_{\rm title},\lambda_{\rm align})$ are 0.05, 0.3, and 0.3. New parameters use the backbone's initialization scale. Each 0.6B \lamp{} run trains for eight epochs with AdamW (weight decay 0.01). A 5\% warm-up precedes cosine decay from peak learning rates of $5\times10^{-5}$ for Qwen3 and $5\times10^{-4}$ for new parameters. Each update uses 128 recommendation examples and 16 alignment items. History and target titles are capped at 16 tokens. A missing title takes the first available category leaf or, if absent, \emph{Unknown}.

\paragraph{Selection and uncertainty}
Full-validation NDCG@10 selects each seed's checkpoint, the MiniOneRec-SFT-AR beams (20, 20, and 100), and $\beval=2$ on Beauty and Sports, with $\beval$ fixed at 4 on Toys. The lower block of Table~\ref{tab:main} reports means and sample standard deviations over seeds 42--44 with a fixed index per domain. Controlled ablations use seed 42 and, for mixtures, $\beval=4$ unless varied. Paired 95\% bootstrap intervals resample test requests, retaining all four models' predictions together for each request in the interaction test.

\paragraph{Timing protocol}
An exclusive RTX~6000D serves one request at a time. The clock starts at CPU collation of a tokenized history and stops when the CPU receives the Top-10, covering transfer to the GPU, inference, and catalog retrieval. We warm up with eight requests and report the p95 across 128 timed requests. Table~\ref{tab:main} uses each model's decoding settings from the accuracy evaluation. The batched serving study fixes $\beval=4$ for \lamp{} across domains and uses the same domain-specific beams for MiniOneRec-SFT-AR.

\subsection{Overall Performance (RQ1)}
\lamp{} leads both matched baselines on all four ranking metrics in every domain (Table~\ref{tab:main}). Its NDCG@10 gains over RPG are 29.9\%, 25.0\%, and 29.9\%. More targets reach the shortlist (Recall@5 up 24.2--30.1\% and Recall@10 up 17.5--30.3\%), and NDCG@5 rises 29.3--37.0\%. The lift in NDCG@5 exceeds the lift in Recall@10 on Toys and Beauty, while the two are comparable on Sports. TIGER leads the published block in NDCG@10 on Toys and Sports, and HSTU on Beauty. Both parallel models in the matched block exceed those published scores. With Qwen3-0.6B-Base and the OPQ32 index held fixed, the grounded four-slot mixture gains 1.161 NDCG@10 points on Toys, 0.921 on Beauty, and 0.656 on Sports over the ungrounded single readout (Table~\ref{tab:mechanism}a, seed 42, $\beval=4$).

On a shared language backbone, \lamp{} also raises early-rank accuracy while shortening the path to a retrieved item. Relative NDCG@10 gains over MiniOneRec-SFT-AR are 127.9\% on Toys, 95.8\% on Beauty, and 89.2\% on Sports. In NDCG@5, the gains range from 102.9\% to 150.7\%. After collision handling, the ratio of distinct MiniOneRec-SFT-AR IDs to catalog items exceeds 99.8\% in each domain. Retrieving an item still requires its prefix to survive every beam step, while \lamp{} scores every complete code before ranking. At the validation-selected beams, MiniOneRec-SFT-AR takes 107.76, 104.93, and 244.34\,ms at p95. \lamp{} returns the Top-10 in 20.52--22.97\,ms, speeding up the same request boundary by $4.7\times$, $4.9\times$, and $11.9\times$.

\subsection{Synergy, Mechanism Ablations, and Tokenizer Generality (RQ2)}
\label{sec:exp-ablation}
\paragraph{Multi-slot and grounding synergy}
In every domain, the gap between the four-slot model and its single-readout reference widens under joint language supervision (Table~\ref{tab:mechanism}a). The single readout uses the final prompt state without grounding and a learned slot with grounding. On Beauty, moving from one to four slots changes NDCG@10 from 4.888 to 4.863 without grounding ($-0.025$ points) and from 5.025 to 5.809 with both language objectives ($+0.784$). The interaction is the difference between these two gains, $+0.809$ points (request-paired 95\% CI $[0.568,1.047]$). The corresponding interactions are $+0.418$ on Toys ($[0.176,0.661]$) and $+0.601$ on Sports ($[0.450,0.757]$). With next-item supervision alone, the slots have few cues to distinguish competing purchases in code space. Titles and item-to-code labels supply those cues to the history representation and the shared classifiers.

\begin{figure*}[t]
\centering
\includegraphics[width=\textwidth]{\lampArxivAssets 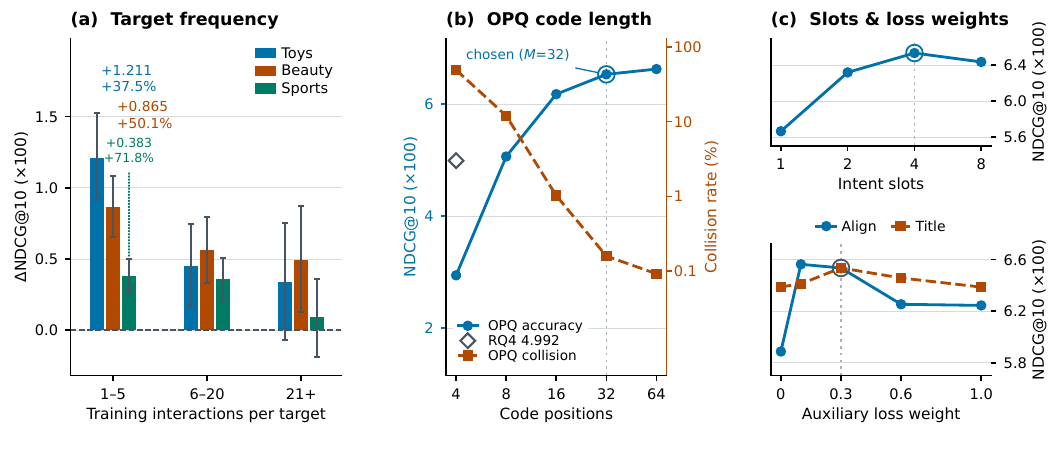}
\caption{Test NDCG@10 ($100\times$, seed 42, $\beta=4$). (a) Adding item alignment to the four-slot, title-supervised model. Bars show absolute gains with request-paired 95\% intervals. Relative gains for items seen 1--5 times are 37.5\%, 50.1\%, 71.8\% on Toys, Beauty, and Sports. (b) Toys accuracy and the percentage of items sharing an OPQ code (log scale). The diamond marks RQ4. (c) Toys slot count and loss weights. Rings mark the default settings in (b,c). A zero weight removes the named objective, leaving the other at 0.3.}
\Description{Target-frequency gains with paired intervals for three datasets; Toys code-length, collision, intent-slot and training-loss weight curves.}
\label{fig:analysis}
\end{figure*}

\paragraph{Readout mechanics and slot specialization}
Under four-slot mixture training, item alignment alone reaches 6.388, 5.604, and 3.234, ahead of title prediction alone at 5.888, 5.193, and 3.158. Using both brings the scores to 6.535, 5.809, and 3.428. The Toys readout controls identify which training signal makes multiple slots effective (Table~\ref{tab:mechanism}b). A single learned slot scores 5.664, while four slots reach 5.823 with CE on the last slot and 5.703 with CE averaged across slots. Holding the four-slot architecture and inference rule fixed isolates the $0.832$-point gain from soft winner-take-all training, which routes more of each target's gradient to the slots that best predict its complete code.

Grounding also separates the predictions of slots receiving the same history. On Beauty and Sports, mean pairwise Top-10 overlap falls from 13.9\% to 8.1\% and from 20.2\% to 8.9\%, respectively. Mean Jensen--Shannon divergence between the slots' 256-way distributions rises from 0.342 to 0.361 and from 0.257 to 0.278, averaging over code positions, all six slot pairs, and test requests. At the population level, the grounded models use 3.82 and 3.79 effective slots, measured as $\exp H(\bar{\mathbf r})$, where $H$ is Shannon entropy and $\bar{\mathbf r}$ averages target responsibilities across all test requests.

\paragraph{Tokenizer generality}
Long flat IDs also work when their construction requires no learned centroids (Table~\ref{tab:mechanism}c). Our SimHash~\cite{simhash} tokenizer draws 256 random Gaussian hyperplanes and packs each group of eight sign bits into a codeword, least-significant bit first. With the same four-slot training recipe, SimHash on raw text vectors scores 6.413 on Toys, 28.5\% above RQ4 (4.992). Applied to the 512-dimensional PCA vectors, the same training-free hash reaches 6.614, slightly above fitted OPQ32 (6.535). Even on the identical four-position residual codebook, \lamp{} scores 4.992 against 3.526 for RQ4-AR with the same title histories and Qwen3 backbone (Section~\ref{sec:exp-scaling}). Across the three 32-position tokenizers, 0.10--0.16\% of catalog items share an ID, compared with 9.05\% under RQ4.

\subsection{Frequency Breakdown and Sensitivity Analysis (RQ3 \& RQ4)}
\label{sec:exp-analysis}
\paragraph{Frequency breakdown of alignment gains}
Item alignment yields its largest gains for targets observed one to five times in training (Figure~\ref{fig:analysis}a). Against the same title-supervised four-slot model, adding alignment raises full-catalog NDCG@10 by $1.211$, $0.865$, and $0.383$ points in this bucket, which covers 27--31\% of test requests in each domain. The paired 95\% intervals are $[0.926,1.523]$ on Toys, $[0.652,1.079]$ on Beauty, and $[0.271,0.501]$ on Sports. Relative gains are 37.5\%, 50.1\%, and 71.8\%. Toys gains the most in absolute terms, moving from 3.229 to 4.440. Absolute gains decline from low to middle to high frequency in every domain. For items seen at least 21 times, the gains narrow to 0.337 points on Toys and 0.095 on Sports, while Beauty retains a 0.495-point gain ($[0.129,0.869]$). Frequent targets already provide repeated sequence-level updates to $W$, whereas item alignment supplies direct supervision where such updates are scarce.

\begin{figure*}[t]
\centering
\includegraphics[width=\textwidth]{\lampArxivAssets 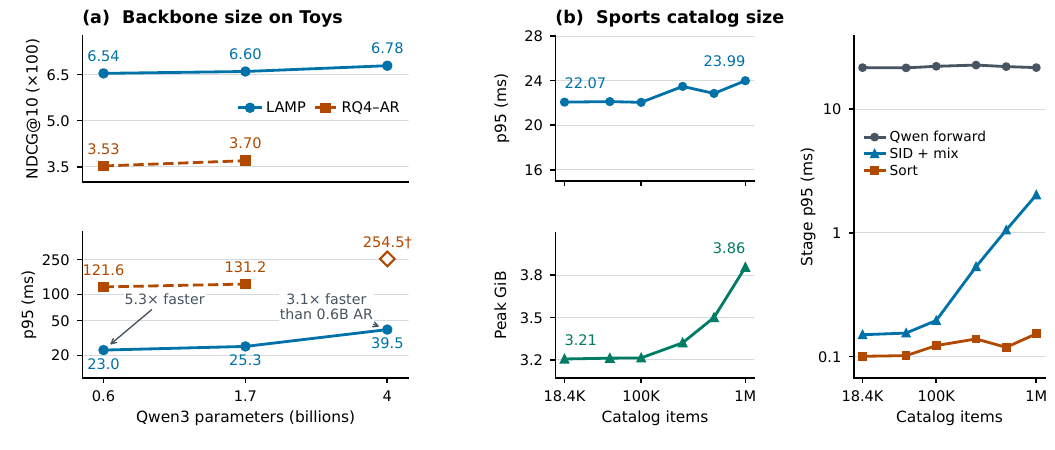}
\caption{Backbone and catalog scaling on an RTX 6000D (batch 1). (a) RQ4--AR uses the same Qwen3 backbone, four autoregressive residual codes, and beam 50. \lamp{} reduces p95 by $5.29\times$ and $5.19\times$ at 0.6B and 1.7B. The open diamond ($\dagger$) times the legal-prefix decoder on a pretrained 4B backbone with an untuned task head. (b) Full-sort retrieval on 128 real Sports prompts. Catalogs scale from 18.4K to 1M items by resampling codes from the 18,357-item catalog. At one million items, the separate SID scoring and sorting stage p95s sum to 2.18 ms.}
\Description{Toys ranking quality and latency by backbone size. Sports full-sort latency, peak reserved GPU memory and stage timings through one million catalog rows.}
\label{fig:scaling}
\end{figure*}

\paragraph{Index resolution at fixed decoding depth}
OPQ gains most from the first increases in code length (Figure~\ref{fig:analysis}b). At $M=4$, 48.99\% of Toys items share an ID, the largest bucket ties 56 items, and NDCG@10 is 2.944, below RQ4's 4.992. At $M=8$, coverage grows from 7,568 to 10,980 distinct tuples and the score reaches 5.067. By $M=16$, only 1.02\% of items share an ID and NDCG@10 has more than doubled to 6.181. Returns taper at our default $M=32$, where 11,924 items occupy 11,911 distinct tuples and NDCG@10 reaches 6.535. Doubling to $M=64$ adds five tuples and 0.094 points (6.629), leaving 0.09\% of items in collisions. It doubles the code-head outputs and catalog lookups per slot while reusing the same backbone states.

\paragraph{Allocating model and supervision capacity}
Four slots perform best among the tested counts on Toys (Figure~\ref{fig:analysis}c). NDCG@10 increases from 5.664 with one slot to 6.320 with two and 6.535 with four, then falls to 6.436 with eight. The second slot gains 0.656 points, whereas moving from four to eight costs 0.099. Each request distributes a total responsibility of one across the slots in Eq.~\eqref{eq:lamp-mix}, so a finer division can leave fewer high-weight updates per slot. For positive weights from 0.1 to 1.0, title supervision yields 6.387--6.535. Alignment yields 6.563 and 6.535 at weights 0.1 and 0.3, falling to 6.254 and 6.245 at 0.6 and 1.0. Increasing this weight gives static item identification a larger share of updates to $W$, while both recommendation and title prediction condition on the purchase history. At inference, across the six tested $\beval$ values from 1 to 12, the test NDCG@10 range stays below 0.06 points for each domain's fixed \lamp{} checkpoint.

\subsection{Backbone Scaling, Catalog Scaling, and Serving Efficiency (RQ5)}
\label{sec:exp-scaling}
% Generated by scripts/phase1_tables.py
\begin{table}[t]
\caption{Qwen3-0.6B-Base on an RTX 6000D. Serving uses \lamp{} $\beta=4$ and MiniOneRec-SFT-AR beams 20/20/100. Bold marks the best completed result per measure. Training includes validation monitoring for \lamp{} and runs through the last checkpoint write for MiniOneRec-SFT-AR.}
\label{tab:systems}
\centering\normalsize
\renewcommand{\arraystretch}{0.94}
\small\setlength{\tabcolsep}{2pt}
\begin{tabular*}{\columnwidth}{@{\extracolsep{\fill}}lrrrrr@{}}
\toprule
\multicolumn{6}{@{}l}{\textit{(a) Serving throughput and peak reserved memory}} \\
Domain & Batch & \multicolumn{2}{c}{\lamp{}} & \multicolumn{2}{c}{MiniOneRec-SFT-AR} \\
\cmidrule(lr){3-4}\cmidrule(lr){5-6}
 &  & req/s & GiB & req/s & GiB \\
\midrule
Toys & 1 & \textbf{45.27} & 3.17 & 10.05 & \textbf{1.80} \\
 & 8 & \textbf{295.82} & \textbf{3.37} & 18.37 & 6.65 \\
 & 32 & \textbf{299.74} & \textbf{4.29} & 16.34 & 24.89 \\
\addlinespace[2pt]
Beauty & 1 & \textbf{45.72} & 3.23 & 10.22 & \textbf{1.79} \\
 & 8 & \textbf{250.02} & \textbf{3.42} & 18.50 & 6.65 \\
 & 32 & \textbf{221.30} & \textbf{4.19} & 16.55 & 24.89 \\
\addlinespace[2pt]
Sports & 1 & \textbf{48.04} & \textbf{3.21} & 5.02 & 5.12 \\
 & 8 & \textbf{274.10} & \textbf{3.42} & 3.50 & 30.84 \\
 & 32 & \textbf{244.36} & \textbf{4.13} & \multicolumn{2}{c}{\textit{OOM}} \\
\midrule
\end{tabular*}\par\nointerlineskip
\begin{tabular*}{\columnwidth}{@{\extracolsep{\fill}}lrrr@{}}
\multicolumn{4}{@{}l}{\textit{(b) Training wall-clock time (min)}} \\
Domain & \lamp{} & MiniOneRec-SFT-AR & Speedup \\
\midrule
Toys & \textbf{111.87} & 370.95 & $\mathbf{3.3\times}$ \\
Beauty & \textbf{160.51} & 353.14 & $\mathbf{2.2\times}$ \\
Sports & \textbf{191.79} & 467.34 & $\mathbf{2.4\times}$ \\
\bottomrule
\end{tabular*}
\end{table}

\paragraph{Backbone size}
\lamp{} gains accuracy as the Toys backbone grows from 0.6B to 1.7B to 4B with a fixed codebook (Figure~\ref{fig:scaling}a). NDCG@10 rises from 6.535 to 6.597 to 6.784, with single-request p95 of 23.0, 25.3, and 39.5\,ms. At beam 50, the fine-tuned 0.6B and 1.7B RQ4-AR controls score 3.526 and 3.698 at 121.6 and 131.2\,ms. At 4B parameters, \lamp{} serves the Top-10 $6.4\times$ faster than the 4B constrained-AR microbenchmark (39.5 vs.\ 254.5\,ms) and $3.1\times$ faster than the fine-tuned 0.6B RQ4-AR control despite using nearly $7\times$ as many parameters. Widening the trained 1.7B AR beam from 20 to 100 more than doubles p95 (94.5 to 216.9\,ms) for a gain of 0.009~NDCG@10~points.

\paragraph{Exact ranking as the catalog grows}
In the Sports catalog-scaling study, history encoding dominates even at one million items (Figure~\ref{fig:scaling}b). We rank the original 18,357 items for each of 128 fixed prompts, then resample item codes to enlarge the catalog. With an exact full sort, end-to-end p95 rises from 22.07 to 23.99\,ms and peak reserved memory from 3.21 to 3.86\,GiB. Backbone p95 remains near 21.7\,ms across both endpoints. At one million items, the p95s for SID lookup and mixture scoring (2.03\,ms) and sorting (0.15\,ms) sum to 2.18\,ms, under 10\% of the 23.99\,ms end-to-end p95.

\paragraph{Serving throughput}
On the same backbone, \lamp{} delivers $4.5\times$--$9.6\times$ the throughput of MiniOneRec-SFT-AR at batch~1. At batches 8 and 32, the ratio reaches $13.4\times$--$78.3\times$ for matched runs that fit in memory (Table~\ref{tab:systems}). \lamp{} serves 250--296 requests/s at batch 8 and 221--300 at batch 32.

\paragraph{Memory and training cost}
\lamp{} serves each batch in a single forward pass, reserving 3.17--4.29\,GiB across all settings. MiniOneRec-SFT-AR caches each surviving beam prefix, reserving 24.89\,GiB on Toys and Beauty at batch 32 and 30.84\,GiB on Sports at batch 8. At batch 32, its 100-wide Sports beam exhausts memory.

{\interlinepenalty=10000
Despite about $2.9\times$ as many optimizer steps, \lamp{} trains $2.2\times$--$3.3\times$ faster (112--192 versus 353--467 minutes, Table~\ref{tab:systems}b). Validation monitoring takes under a minute per run with \lamp{} and 33--44 minutes with beam search.\par}

\section{Related Work}
\label{sec:related}

\paragraph{Sequential and multi-interest recommendation}
Sequential recommendation models how earlier choices shape the next one, from personalized Markov transitions~\cite{fpmc} to recurrent, convolutional, and attention-based encoders~\cite{gru4rec,caser,sasrec,bert4rec,s3rec} and large-scale sequential transducers~\cite{hstu}. UniSRec and Recformer use item descriptions to transfer what a sequence encoder learns across catalogs~\cite{unisrec,recformer}. DIN and DIEN use a candidate to identify relevant past behavior~\cite{din,dien}, while MIND and ComiRec retrieve candidates with distinct interest vectors drawn from the same history~\cite{mind,comirec}. ColBERT, ColBERTv2, and WARP likewise retain multiple query and document vectors until matching time~\cite{colbert,colbertv2,warp}. These methods score similarities between dense vectors. In discrete ID space, \lamp{} scores each candidate's complete code under competing interpretations of a history in one backbone pass.

\paragraph{Generative recommendation and semantic IDs}
DSI and NCI retrieve documents by generating identifiers from queries~\cite{dsi,nci}, and GenRet learns the document tokenizer together with the retrieval model~\cite{genret}. TIGER applies generative retrieval to recommendation by decoding the next item's residual codes in sequence~\cite{tiger}. LETTER incorporates collaborative structure into the tokenizer, while ETEGRec and DIGER train identifiers to support the recommendation objective~\cite{letter,etegrec,diger}. Recent work reduces serial decoding through parallel code prediction or multi-aspect set prediction~\cite{rpg,setrec,flash} and transfers a sequential teacher's predictions to parallel decoders in item or code space~\cite{kdparallel}. Predicting multiple aspects of one item and representing competing next purchases are distinct tasks. A unimodal marginal factorization can mix codewords from different plausible items and produce ghost ties. \lamp{} preserves intra-item codeword coupling by scoring the complete ID under each intent before pooling across intents.

\paragraph{LLM-based and language-aligned recommendation}
P5 casts recommendation tasks as text generation, and TALLRec tunes a language model on recommendation instructions~\cite{p5,tallrec}. LLaRA pairs collaborative embeddings with item text in the prompt~\cite{llara}, while LC-Rec aligns collaborative identifiers with language through instruction tuning~\cite{lcrec}. Once a language model can generate item IDs, its training objective also shapes which candidates survive retrieval. MiniOneRec combines supervised fine-tuning with reinforcement learning~\cite{minionerec}, and a study of RL post-training finds that gains at the top of a ranked list can accompany lower recall at larger cutoffs~\cite{rlretrieval}. In \lamp{}, predicting the next title grounds the history slots in language, and predicting an item's ID from its text trains their shared code head. Both objectives are optimized jointly with next-item prediction. The resulting model ranks the catalog directly, with a single backbone pass and no separate reranker.

\section{Conclusion}
\label{sec:conclusion}
{\interlinepenalty=10000
We presented \lamp{}, a single-pass generative recommender that preserves intra-item codeword coupling by scoring each candidate's complete semantic ID under multiple intent slots before pooling. Pairing this mixture decoder with training-only bidirectional language grounding anchors both the history slots and the shared code head at zero inference cost. Across three Amazon domains, the four-slot advantage over the corresponding single readout grows with both language objectives, and item-text alignment yields its largest gains on low-frequency items ($1$--$5$ training interactions); the same decoder also transfers to SimHash IDs built without tokenizer training. Under a unified full-catalog protocol, \lamp{} raises three-seed NDCG@10 by 25.0--29.9\% over RPG; on the same Qwen3-0.6B backbone, it outperforms MiniOneRec-SFT-AR by 89.2--127.9\% while cutting single-request p95 latency by $4.7\times$--$11.9\times$ and increasing batched throughput by $13.4\times$--$78.3\times$. On Toys, accuracy improves from 0.6B to 4B parameters. The 0.6B model ranks a resampled one-million-item catalog exactly in 24.0\,ms at p95.\par}

\paragraph{Future work}
Prompt compression and prefix caching could reduce history encoding time, which accounts for about 21.7\,ms at p95 on Sports. For industrial catalogs with tens or hundreds of millions of items, a graph index over product codes~\cite{rpg} could retrieve a shortlist for exact multi-slot reranking. A further direction is to extend language-to-SID alignment to retrieve new products before their first interaction.

\label{lamp:body-end}
\setlength{\bibsep}{2.6pt plus 1pt minus 1pt}
\RaggedRight
\bibliographystyle{ACM-Reference-Format}
\bibliography{references}
\end{document}